# Evaluation of SystemC Modelling of Reconfigurable Embedded Systems

Tero Rissa
Department of Computing
Imperial College London, UK
tpr@doc.ic.ac.uk

Adam Donlin
Xilinx Research Labs
Xilinx Inc.
adam.donlin@xilinx.com

Wayne Luk
Department of Computing
Imperial College London, UK
wl@doc.ic.ac.uk

**Abstract**

*This paper evaluates the use of pin and cycle accurate SystemC models for embedded system design exploration and early software development. The target system is MicroBlaze VanillaNet Platform running MicroBlaze uClinux operating system. The paper compares Register Transfer Level (RTL) Hardware Description Language (HDL) simulation speed to the simulation speed of several different SystemC models. It is shown that simulation speed of pin and cycle accurate models can go up to 150 kHz, compared to 100 Hz range of HDL simulation. Furthermore, utilising techniques that temporarily compromise cycle accuracy, effective simulation speed of up to 500 kHz can be obtained.*

## 1 Introduction

Introduction of embedded processors into Field Programmable Gate Arrays (FPGAs) exposes mainstream designers to a new field of CAD: hardware and software development of application specific embedded systems. This creates a conflict between two verification worlds. Software developers are accustomed to high-speed Instruction Set Simulators (ISS) with full visibility and controllability to the system, whereas hardware engineers use Hardware Description Language (HDL) simulators. Currently the latter is the only option provided for complete system simulation, which is clearly too time consuming as illustrated later in this paper.

This paper examines the usefulness of SystemC modelling to alleviate the gap between embedded software and hardware verification methods. Previous related work include, amongst others, SystemC modelling of an IP forwarding chip [1], AMBA bus architecture [2], multiprocessor SoC platforms [3], and virtual in-circuit emulation [4]. This paper presents several different SystemC models of a complex embedded system executing uClinux operating system. It is shown that there can be 8-fold difference in the simulation time depending on the used SystemC model.

The rest of the paper is organised as follows. Section 2 introduces the target system and the used SystemC tools. Section 3 briefly looks in to the HDL RTL simulation of the system. Sections 4 and 5 presents the cycle and non-cycle accurate SystemC models and simulation speed results. Section 6 highlights some of the drawbacks and future improvement possibilities with SystemC modelling. Finally section 7 concludes the paper.

## 2 Target System

The target system is uClinux [5] running on MicroBlaze VanillaNet [6] platform depicted in Fig. 1. Both the hardware version of the platform and the Linux port are created by John Williams [7, 8]. Publicly available third party platform and application was chosen to facilitate comparisons with our models. uClinux is a derivative of Linux 2.0 kernel intended for microcontrollers without Memory Management Units (MMUs). The MB VanillaNet platform is targeted to Insight/Memec V2MB1000 Virtex2 evaluation board and the range of peripherals reflects this.

All the simulations are carried out using IBM IntelliStation Z Pro with 3.06 GHz Intel Xeon processor with 2.5 GB RAM memory, running Red Hat 8 operating system with Linux 2.4.18-14 kernel. Each SystemC simulation result is an average of 50 data points: 10 different phases over 5 executions of the Linux boot sequence. From obvious reasons the RTL HDL simulation results are not from Linux boot sequence, but from a simpler program execution.

### 2.1 SystemC Tools

All the SystemC development tools used within this project are Free Open Source Software (FOSS). SystemC class library, including the source code is free and available to the public via SystemC portal [9]. In addition to standard Linux C++ development and shell tools, GTKWave waveform viewer and Data Display Debugger (DDD) were used.

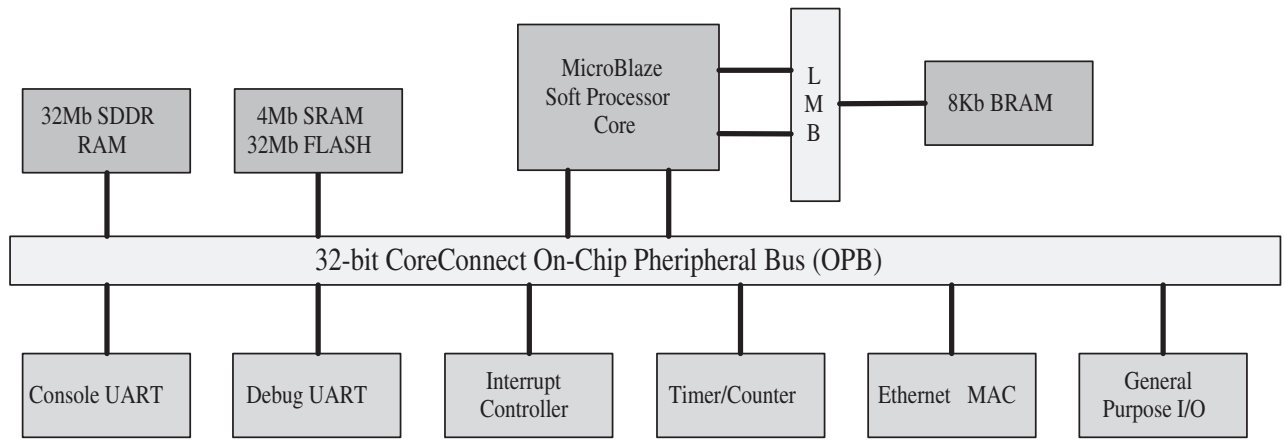

**Figure 1. Target platform: MicroBlaze VanillaNet**

The availability of full set of FOSS development tools to carry out project with this magnitude indicates that the financial threshold for adaptation of SystemC form the tool cost point of view is negligible.

## 3 RTL HDL Simulation

The RTL HDL simulation is carried out in Mentor Graphics ModelSim SE 6.0 without any trace waveform. The used HDL model is generated by Xilinx Embedded Development Kit (EDK). As seen from Fig. 2, which contains the results from all the presented models, the HDL simulation can be run with the speed of 167 Hz. With this simulation speed, it would take 1 month and 15 days to complete uClinux boot sequence.

## 4 Pin and Cycle Accurate Models

The SystemC models presented in this section are pin and cycle accurate models of the target system components; peripherals, buses and the MicroBlaze. This means that the all the signals between the components that are present in the RTL HDL model are also present in the SystemC models. Furthermore, the waveforms of these signals in these models are identical. The difference is that pin and cycle accurate model does not model all the signals and register transfers inside the components. For example, multi cycle operation can be carried out in zero simulation time and then the result delayed for required amount of cycles. In general this means that the components' internal description can be done using standard C++, whilst SystemC is used for the component interface. Naturally, SystemC must be used also for concurrent elements to maintain the cycle accuracy within the interface. A notably large component is the Xilinx MicroBlaze ISS, which is standard C++ implementation wrapped in SystemC module.

The UART models connects to the pseudo terminal (PTY) interface on the underlying Linux operating system. By using this interface, it is possible to connect to the model using standard terminal software, such as `minicom`. The SystemC model of Ethernet MAC is a proxy that implements only the OPB interface and peripheral control registers.

### 4.1 Initial SystemC Model

In the initial SystemC model, `sc_[in|out]_rv` ports are used in all the system components and `sc_signal_rv` signals are used to connect the ports. Within modules, native C++ types are used, whenever possible. The main reason of using resolved signal and port types is to enable HDL co-simulation in ModelSim. It is well known that these data types are slower to simulate than unresolved ones. However, the simulation speed of this type of model is already 61 kHz – 360 times faster than RTL HDL simulation.

### 4.2 Native C++ Data Types

The fist optimised model utilises native C++ data types for signals between system components. When using these data types, HDL co-simulation is no longer possible. In addition, multiple drivers of a same signal are no longer detected. The use of this optimisation is facilitated by signal declaration and manipulation macros and inline functions. These constructs makes is possible to turn the optimisation on and off during compilation time without changes to the source code of the models.

As it can be seen from Fig. 2 this optimisation provides 132% speed improvement compared to the previous model, yielding simulation speed of 141.7 kHz.

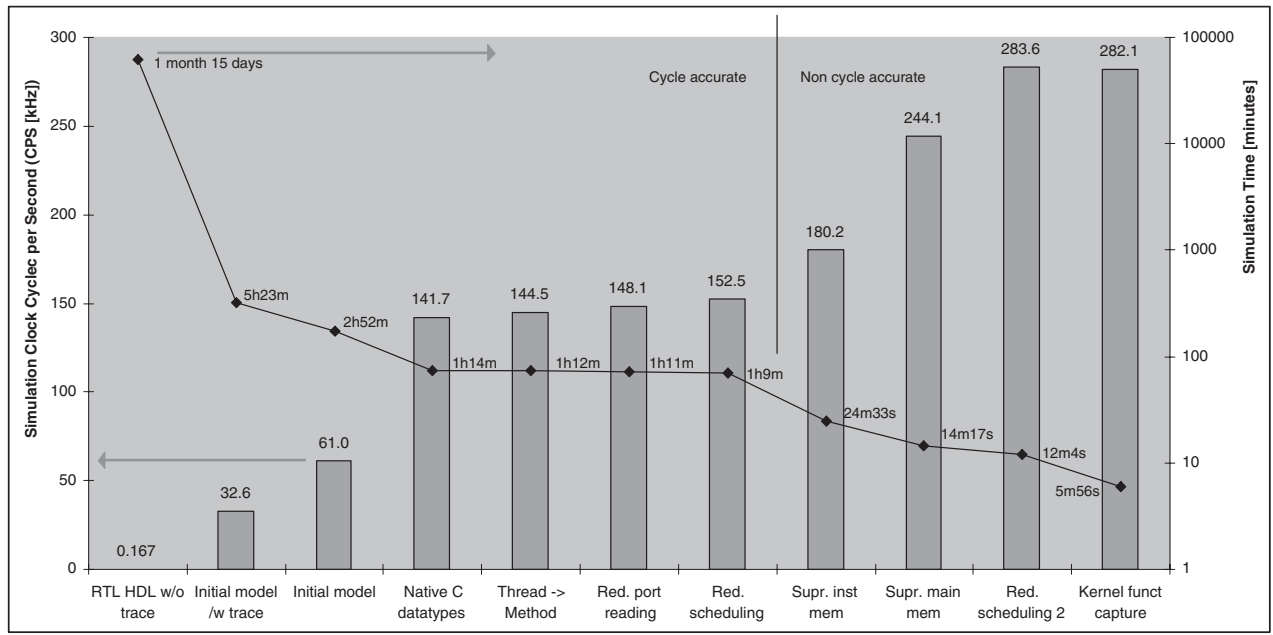

**Figure 2. Results. Bars for simulation speed (CPS/[kHz]) with values on the left and line plot for simulation time [min] with values on the right.**

### 4.3 Threads vs. Methods

`SC_THREAD` is a SystemC process that can have multicycle behaviour. When multicycle behaviour is not required, more lightweight process – `SC_METHOD` – can be used. Methods are faster to simulate than threads due to reduced scheduling complexity. In this model, 3 out of 17 processes in total are changed from threads to methods. It can be seen that this yields modest 2% speed improvement.

### 4.4 Reduced Port Reading

In SystemC, ports are implemented as C++ objects. Every port operation, e.g., read and write, will inflict a chain of function calls due to request-update nature of the concurrency kernel. Listing 1 illustrates how port reading can be reduced, when port value is required several times within single process execution. The first code snippet uses the value of `input_X` two times if the value is not 2. By utilising local variable to store the input value, only one reading of the port is required. Naturally, this technique can only be utilised, when reading of the port is not blocking operation and does not consume port item, as can be the case for example with `sc_fifo`.

In this model, 6 input port reads occurring every cycle were reduced to 3. This yields 2.5% speed improvement. As in this model there are about 70 port read operations every cycle, slightly depending on phase of execution, reducing port reading has potential for notable speed improvement. Especially, it can be noted that hardware description style to check of reset condition on every cycle is not the most efficient way of writing SystemC code. Instead, the variables should be initialised for example in the constructors, or global variable, not a port, used for reset detection.

**Listing 1. Reduced port reading**

```
// Multiple port reads
void input_method() {
  if(input_x.read() != 2) {
    z = input_x.read() + input_y.read();
  }
}

// Reduced port reads
void input_method() {
  unsigned int local_X = input_x.read();
  if(local_X != 2) {
    z = local_X + input_y.read();
  }
}
```

**Listing 2. Reduced port reading**

```
// Separate threads for operations
void thread1() {
  while(1) {
    z = x+y;
    wait();
  }
}
void thread2() {
  while(1) {
    answer = z + 42;
    wait();
  }
}

// Separate functions for operations
void combined_thread() {
  while(1) {
    do_function2();
    do_function1();
    wait();
  }
}
void do_function1() {
  z = x+y;
}
void do_function2() {
  answer = z + 42;
}
```

## 4.5 Reduced Scheduling

In pin and cycle accurate modelling vast majority of SystemC processes, i.e., threads and methods, are scheduled every clock cycle. The models presented in this section reduces scheduling with two techniques explained in sections 4.5.1 and 4.5.2. These techniques has potential to yield speed improvement through two factors: 1) reduce the computation in SystemC scheduling engine and 2) reduce the overall computation of a process. The first factor is present in both techniques, the second one only in the latter.

### 4.5.1 Combination of Concurrent Elements

In RTL HDL description it is common practise to declare separate process for every state of processing in order to make the code more understandable and modular. In SystemC, although separate processes – methods or threads – have identical sensitivity, every process is scheduled separately. As SystemC is based on C++ it is natural to perform the computation in functions that are called from processes instead of performing the computation inside process directly. The difference is that the functions can be called from a single process without compromising the readability and modularity of the code. As these models are not targeted for hardware implementation, synthesisability of the function calls is not an issue, unlike in RTL HDL models.

Listing 2 demonstrates this optimisation in practice. The upper code snippet has two separate processes, in this case threads, for computation. In the lower snippet, the computation is done inside functions that are called from `combined_thread()`. It is good to note the order of function calls. `do_function2()` must be called before `do_function1()` in order to achieve identical behaviour regardless of whether SystemC signals or native C++ data types were used for variables. The computation is usually rather more complicated than in this example and identifying the correct sequence of operations becomes increasingly difficult as complexity increases. This limits the applicability of the optimisation.

In this model, 3 synchronous single cycle threads are combined to a single thread. As can be seen from Fig. 2 this yields 3% speed improvement. This illustrates that the scheduling in SystemC kernel is notable part of the execution. In addition, asynchronous processes that are potentially scheduled several times within a single clock cycle, should be avoided whenever possible.

### 4.5.2 Multicycle Sleep of Processes

In some cases, processes (methods and threads) do not need to be invoked every clock cycle in order to achieve the desired functionality. For example in the target system, the underlying PTY interface of the UART model is capable of receiving characters much faster than our model is able to fill the transmission buffer. Therefore, instead of scheduling the transmission process to execute every cycle, the process can sleep for several cycles between executions. In SystemC, this can be done using `wait(time, sc_time_unit)` and `next_trigger(time, sc_time_unit)` methods for threads and methods, respectively.

This optimisation can be applied in relatively rare occasions and the benefit greatly depends on amount of computation done in each time the process is scheduled. In the case of UART interface module, the transmission will cause system calls, that have relatively long and host system load dependent execution times. In order to reduce system calls, and thus speed distortion of the results, this optimisation is utilised in all of the presented models.

### 4.6 Summary of Pin and Cycle Accurate Models

Unoptimised pin and cycle accurate SystemC model provided simulation speed of 61 kHz, which already provides around 360 times speedup compared to RTL HDL simulation. Mainly by utilising native C++ data types instead of multi-valued SystemC data types, 152.5 kHz simulation speed was obtained. The last three optimisation techniques had relatively small impact, 7.6% altogether, to the overall simulation speed. On the other hand, utilisation of these optimisations is simple and the tradeoffs are minor. Therefore, these techniques should be considered more as good modelling style guidelines than optimisation methods.

## 5 Non-Cycle Accurate Models

The models presented in this section uses optimisation techniques that does not preserve cycle accuracy. All the optimisation can be turned on and off during run time of the simulation. This makes it possible to quickly simulate section of the execution that is already known to work properly to a section that requires cycle accuracy. In general, it is possible to miss some errors in the system, when these optimisations are utilised.

### 5.1 Instruction Memory Activity Suppression

Vast majority of the OPB bus activity is MicroBlaze instruction fetches from SDDR RAM. In this model, the instruction fetches are suppressed by a new module called memory dispatcher. This module can directly access the memory models inside the peripherals. The benefit of providing the instructions via memory dispatcher are twofold: there are no longer arbitration conflicts between MicroBlaze data and instruction side OPB and even more importantly, it takes only one cycle to fetch the instruction instead of the minimum of three. As can be seen from Fig 2, the improvement in CPI is around 35%, whereas the execution time goes down 64% – from 1 hour 9 minutes to 24 minutes.

Clearly, this model compromises cycle accuracy in several ways. However, the operation of the memory dispatcher can be turned on and off at run-time.

### 5.2 Main Memory Activity Suppression

With this model, the whole main memory peripheral – SDDR RAM – is handled by the memory dispatcher. Again, the speedup is due to two factors. As in previous model, the number of cycles required for memory operation is reduced. Furthermore, as the memory peripheral is completely handled by the memory dispatcher, is no longer necessary to schedule the memory peripheral. Therefore, when the memory dispatcher is enabled, memory peripheral's attachment to the OPB can be removed.

### 5.3 Further Reduction of Scheduling

In the two previous models, the main speedup was due to simplification of frequently occurring peripheral transactions. In the target system, the uClinux has very small number of transaction with FLASH memory, GPIO, and Ethernet MAC peripherals. However, these peripherals' address decode unit is scheduled every clock cycle. This unnecessary scheduling and address decoding activity can be suppressed by explicitly accessing the peripheral only when the address is within the correct range.

As can be seen from Figure 2 this optimisation reduces the boot time to 12 minutes, providing additional 15% speedup. The danger of this optimisation is that it is no longer possible to detect some corner case bugs in the system, as for example peripheral taking over the bus when not allowed.

### 5.4 Interception of Kernel Functions

Final model utilises interception of two low level C-library functions. The Linux boot execution spends 52% on two functions: *memset* and *memcpy*. The former sets a memory region to a certain byte-value and the latter performs byte-wise copy of non-overlapping memory region to another.

The interception of the function is carried out by the SystemC MicroBlaze ISS wrapper. First, the wrapper detects jump to either function and reads function parameters from the MicroBlaze registers. Then the function is executed in native C++ on the host computer in zero simulation time. Finally, the wrapper modifies the ISS registers to have the same values than after normal function execution. In both of the cases only one instruction – the loop check branch – is different compared to the normal execution.

With this optimisation, the CPI is actually lower, as illustrated by Fig 2, because of extra computation required to perform the captures. As roughly half of the instructions are executed in zero time, it is not surprising that the boot-up execution time is also halved from 12 minutes to 6 minutes. Effectively this boot time corresponds to simulation speed of 578 kHz. This optimisation is very application specific and does not only compromise cycle accuracy, but also the ISS execution statistics will be distorted.

### 5.5 Summary of Non-Cycle Accurate Models

The optimisation techniques yielding non-cycle accurate models provide up to 10 000 times speedup compared to RTL HDL simulation. All the techniques considered within this study can be turned on and off during the simulation. This makes it possible to quickly simulate portions of the code that are already known to function correctly and then return to cycle-accurate model for portions that require more detailed examination. However, it must be noted that although the optimisation can be turned on and off, the system will not be in exactly identical state compared to fully cycle accurate simulation. For example, interrupts will occur in different phase of the execution, resulting different program counter traces. In general, this is a problem only in most pathological cases as for example interrupts should function correctly regardless of the phase of execution.

## 6 SystemC Drawbacks and Solutions

This section outlines some of the drawbacks of SystemC modelling experienced within this project.

The major drawback for software development is that standard software development tools are debugging the software *of* the model, not the software running *on* the model. For example, the code that is being traced is SystemC source code, not the source code of the uClinux. One possible solution to this problem would be to provide debugger interface, for example to GDB, from the SystemC model.

There is no linter available for SystemC. In other words, the SystemC semantics that is build on top of C++ syntax, is not checked within the compilation process. The effect of this is that illegal semantics that are syntactically correct will not produce compiler errors or warnings. In these occasions, the programs will cause a run-time error, which are usually harder to locate than compile-time errors. In addition, standard C++ compiler will produce undecipherable error messages, when the illegal use of SystemC semantics yields and syntactical error within the SystemC library.

Interaction with other software environments and native C/C++ and SystemC can be troublesome. This is due to two reasons. 1) `main()` function of the SystemC executable of within the SystemC kernel and thus not easily modifiable by the programmer. This implies that SystemC kernel is the top-level entity in the system and all the functionality must be executed within SystemC components. 2) There is no standard way of creating SystemC events from non-SystemC sources. Significance of this is that SystemC processes cannot be made sensitive to anything that is not implemented with SystemC channel types. For example in the case presented in the paper, it would have been useful to be able to be sensitive to the MicroBlaze ISS output variables, which are implemented in native C++.

## 7 Conclusion

Pin and cycle accurate SystemC modelling provides high-speed alternative to RTL HDL simulation of reconfigurable embedded systems. This paper demonstrated the usability and speed of SystemC model simulation by booting up uClinux on SystemC model of MicroBlaze VanillaNet platform. The speed of cycle accurate SystemC models range from 60 kHz to 150 kHz. Models that have capability of turning cycle accuracy on and off during the simulation can achieve simulation speed of up to 280 kHz. However, due to techniques used, the effective simulation speed can be up to 578 kHz. The speedup compared to RTL HDL simulation ranges from 360x to 10 000x, depending on the used SystemC model. Due to higher simulation speed, SystemC modelling facilitates early embedded software development and enables rapid and easy architectural exploration.